# Fresnel Magnetic Imaging of Ultrasmall Skyrmion Lattices

Yongsen Zhang, Wei Liu, Meng Shi, Yaodong Wu*, Jialiang Jiang, Sheng Qiu, Huanhuan Zhang, Hui Han, Mingliang Tian, Haifeng Du, Shouguo Wang* and Jin Tang*

Y. Zhang, S. Wang

Anhui Provincial Key Laboratory of Magnetic Functional Materials and Devices, School of Materials Science and Engineering, Anhui University, Hefei 230601, China

Email: sgwang@ahu.edu.cn

W. Liu, H. Han

Institutes of Physical Science and Information Technology, Anhui University, Hefei 230601, China

M. Shi, S. Qiu, M. Tian, H. Du

Anhui Provincial Key Laboratory of Low-Energy Quantum Materials and Devices, High Magnetic Field Laboratory, HFIPS, Chinese Academy of Sciences, Hefei, Anhui 230031, China

Y. Wu

School of Physics and Materials Engineering, Hefei Normal University, Hefei, 230601, China

Email: wuyaodong@hfnu.edu.cn

J. Jiang, H. Zhang, J. Tang

State Key Laboratory of Opto-Electronic Information Acquisition and ProtectionTechnology, School of Physics, Anhui University, Hefei, 230601, China

Email: jintang@ahu.edu.cn

**Abstract**

Magnetic skyrmions with ultrasmall nanometric dimensions hold significant promise for next-generation high-density spintronic devices. Direct real-space imaging of these topological spin textures is critical for elucidating their emergent properties at the nanoscale. Here, we present Lorentz transmission electron microscopy studies of nanometric skyrmion lattices in B20-structured $Mn_{0.5}Fe_{0.5}Ge$ crystals using Fresnel mode. According to conventional chiral discrimination methods relying on static bright-dark contrast, we demonstrate an abnormal periodic chiral-reversal phenomenon retrieved through the transport of intensity equation analysis of defocus-dependent Fresnel images. Through systematic off-axis electron holography experiments and numerical simulations, we attribute these chiral misinterpretations to the sinusoidal modulation mechanism of the contrast transfer functionthat correlates with both defocus values and skyrmion dimensions. Our findings establish quantitative limitations of conventional Fresnel contrast analysis for ultrasmall skyrmions while revealing fundamental insights into defocus-mediated phase-to-intensity conversion processes in nanoscale magnetic imaging.

# 1. Introduction

Magnetic skyrmions, characterized by their small size[1-3], low energy consumption[4, 5], rapid dynamic response[6, 7], and robust stability[8, 9], are pivotal for representing data bits and could revolutionize information processing and storage[10-14]. Recent research has predominantly focused on larger-scale structures ( > 50 nm)[15-17], while studies addressing skyrmions at the 10-nm level remain relatively limited. Reducing the size of skyrmions allows for more data bits to be accommodated within the same physical space, thereby offering the possibility of achieving higher-density information storage and promoting the development of more compact and efficient storage devices[18-20].

The evolving paradigm of magnetic characterization methodologies exhibits complementary strengths and intrinsic limitations: Spin-polarized microscopy[21] achieves sub-nanoscale magnetic resolution but demands atomically flat surfaces and extremely low temperatures. Neutron diffraction probes reciprocal-space representations of macroscopic magnetic ordering yet lacks spatial specificity for resolving isolated nanoscale spin textures[1]. In contrast, Lorentz-TEM operated in Fresnel mode leverages electron wave phase modulation to achieve direct real-space mapping of magnetic flux distributions. Its advantages are particularly manifested in three key dimensions: (1) Sub-nanometer resolution for resolving helical/vortex-type spin topologies[22, 23]; (2) Chirality visualization for conventionally-sized skyrmions can be achieved by leveraging the contrast differences between bright and dark regions produced via objective lens defocusing, enabling direct differentiation between clockwise and counterclockwise magnetic moment distributions[24], [25]; and (3) compatibility with in-situ dynamic interrogation of magnetic evolution processes[26-30]. Synergistic integration with off-axis electron holography further minimizes delocalization artifacts, establishing this multimodal approach as a powerful toolkit for nanoscale magnetic structure elucidation[31-33]. However, Fresnel magnetic imaging has some limitations: it cannot distinguish Néel-type spin textures from uniform magnetizations, and the arbitrary filter parameters used in Transport of Intensity Equation (TIE) analysis risk introducing artificial spin textures[34-38]. A critical question warranting in-depth investigation is whether the established criteria for identifying conventionally-sized skyrmions–such as the direct chirality determination method based on Fresnel contrast–remain valid when applied to their ultrasmall counterparts. This fundamental issue of size-dependent validity directly impacts our mechanistic understanding of magnetic imaging principles. Comprehensive elucidation of the physical mechanisms governing Fresnel magnetic imaging, particularly its inherent scale correlation with spin textures, proves essential for unlocking the full potential of ultrasmall skyrmions in advanced technological applications.

In this study, we implement an integrated methodology combining Fresnel mode, off-axis electron holography, and micromagnetic simulations to investigate 10-nm-scale skyrmion lattices in $Mn_{0.5}Fe_{0.5}Ge$ thin lamella[23], [39], [40]. Previous studies have shown the experimental imaging of 10-nm skyrmions in MnGe and $Mn_{1-x}Fe_xGe$ systems[23], [41]. In contrast, Our systematic experiments uncover a periodic contrast inversion phenomenon under varying defocus conditions, with inversion periodicity exhibiting a strong correlation with skyrmion dimensions. Our work clarifies that relying solely on black-and-white contrast for determining the chirality of ultra-small skyrmion lattice in Fresnel-mode imaging can lead to misinterpretation. Furthermore, we explored this characterization method to elucidate the origins of the periodic contrast variations observed in nanometric skyrmion lattice as a function of defocus amount in the Fresnel mode.

## 2. Results and Discussion

### 2.1 Field-driven magnetic evolution and magnetic phase diagram in $Mn_{0.5}Fe_{0.5}Ge$

The $Mn_{0.5}Fe_{0.5}Ge$ crystal investigated in this study exhibits a B20 structure[23], belonging to the non-centrosymmetric space group P213, with a Curie temperature $T_C$ of approximately 170 K (see Figure. S1). In this chiral lattice helimagnet, the spin system is modeled by the following effective Hamiltonian[42]:

$$H = \int d\mathbf{r}[\frac{J}{2}(\nabla \mathbf{M})^2 + \alpha \mathbf{M} \cdot (\nabla \times \mathbf{M})] \quad (1)$$

Where $\mathbf{M}$ denotes the spatially varying magnetization, $J$ represents the ferromagnetic exchange interaction, $\alpha$ is the DMI constant, and $\mathbf{r}$ is the three-dimensional position vector. In the ground state, an appropriate helical magnetic structure is stabilized. Within this model, the magnitude of the wavevector $\mathbf{q}$ is proportional to $\alpha/J$. Due to the asymmetric DMI, the chirality of the spin helical structure (hereafter referred to as magnetic helicity) depends on the sign of $\alpha$. It is known that the sign of $\alpha$ is determined by both the crystal chirality and the sign of the spin-orbit coupling. When the propagation vector $\mathbf{q}$ is parallel (or antiparallel) to $M_1 \times M_2$ (where $M_1$ and $M_2$ are the magnetic moments aligned along the $q$ direction), we define it as a right-handed (or left-handed) helix.

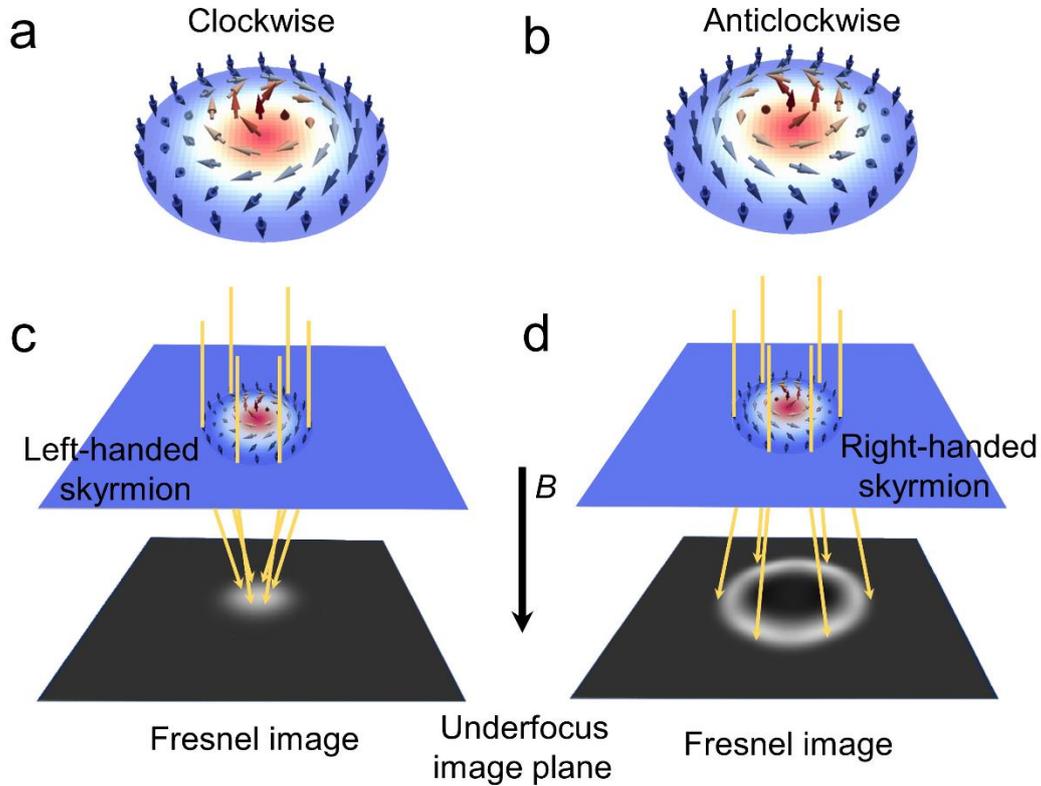

**Figure 1.** (a) and (b) Schematic spin configuration of skyrmions when the external magnetic field *B* is applied downwards. (c) and (d) Schematic illustrations of Lorentz-TEM used in defocused conditions to observe skyrmions and determine their helicity. The orange lines represent the electron beam.

Upon applying an external magnetic field *B* perpendicular to the lamella, skyrmions emerge in the helimagnet. Under these conditions, two configurations of skyrmions can be stabilized: the magnetic moments rotate clockwise (CW) or anticlockwise (ACW) in the plane, corresponding to left- and right-handedness, as shown in **Figure 1**a and 1b. The chirality of skyrmions is not only a cornerstone of their topological stability but also a critical parameter in regulating their physical properties, such as mobility and response to external fields[43], as well as in enabling functional applications like data storage and logic devices[44]. A profound understanding of chirality will drive advancements in topological spintronics, particularly in achieving high-density information storage technologies. Lorentz-TEM is a powerful tool for visualizing the real-space magnetization distribution of topological spin textures. The incident electron beam is deflected by the Lorentz force due to the local in-plane magnetic induction within the sample, and the spatial variation of the in-plane magnetization results in convergence (bright contrast) or divergence (dark contrast) on the defocused image planes. Using this method, helical magnetic structures appear as stripes. When the objective-lens current of the TEM generates an applied magnetic field *B*, Lorentz-TEM images contain sufficient information to reflect the helicity of the skyrmions. As illustrated in Figure 1c and 1d, an in-plane CW (ACW) magnetic moment configuration acts as a convex (concave) lens, forming bright (dark) spots on the defocused image planes[45-47]. Consequently, in defocused images, the position and helicity of conventionally-sized skyrmions are simultaneously visualized as spotty images and their contrast.

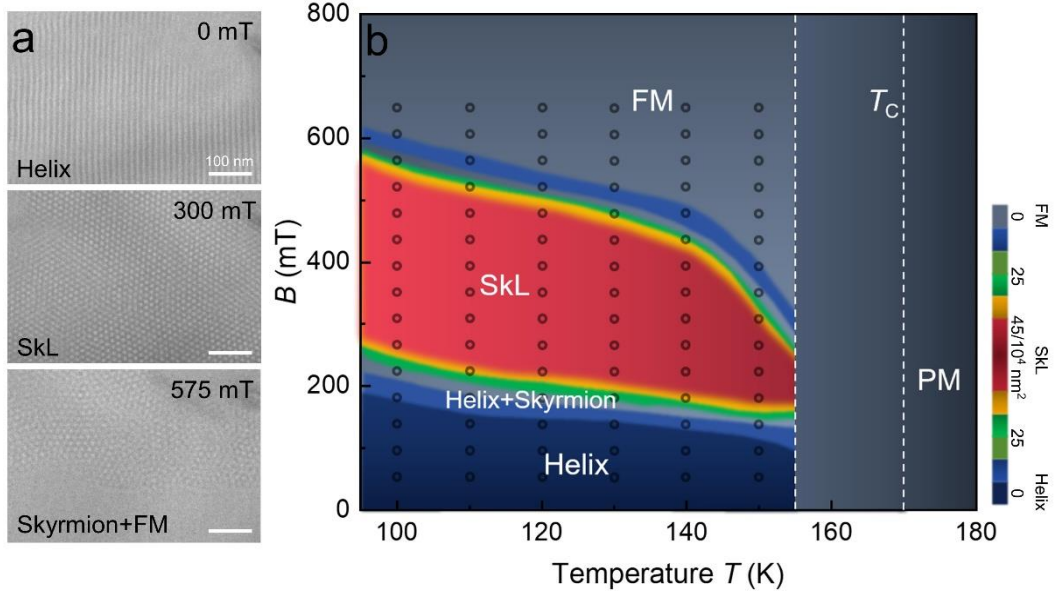

**Figure 2.** (a) The evolution of magnetic structures under out-of-plane magnetic fields at 95 K in a 150 nm-thick $Mn_{0.5}Fe_{0.5}Ge$ lamella. Defocus, 20 μm. (b) Magnetic phase diagram of magnetic skyrmion as a function of temperature and magnetic field. SkL, FM, and PM represent skyrmion lattice, ferromagnetic, and paramagnetic states, respectively. The color bar in the phase diagram indicates the skyrmion density per $10^4$ $nm^2$.

We first examined the magnetic structures within a 150 nm-thick $Mn_{0.5}Fe_{0.5}Ge$ lamella using Lorentz-TEM. As shown in **Figure 2**a, at 95 K and in the absence of an external magnetic field, we observed a typical ground-state helical magnetic domain structure. As the perpendicular magnetic field was gradually increased to 300 mT, the helical domains rapidly transformed into skyrmions, which further evolved into a compactly packed skyrmion lattice configuration. The size of the skyrmions ($\lambda_{sk}$) was approximately 15 nm. Upon further increasing the magnetic field to 575 mT, the skyrmions gradually annihilated and eventually transitioned into a ferromagnetic (FM) state. By systematically varying the magnetic field at different temperatures, we constructed a magnetic phase diagram that describes the evolution of magnetic domains (see Figure 2b). Below the $T_C$, small-sized skyrmions in $Mn_{0.5}Fe_{0.5}Ge$ persist primarily in a lattice form over a wide range of magnetic fields.

**2.2 Fresnel imaging of the skyrmion lattice in $Mn_{0.5}Fe_{0.5}Ge$**

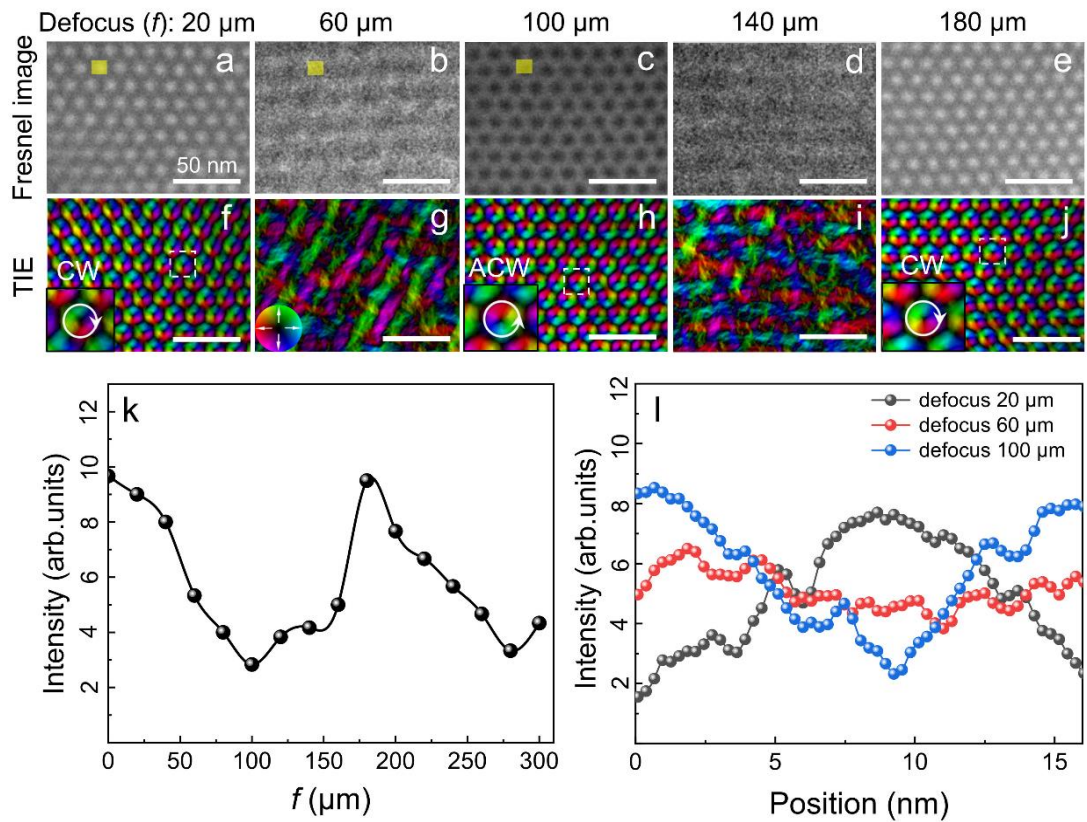

**Figure 3.** (a)–(e) Fresnel images obtained at defocus values of 20 μm, 60 μm, 100 μm, 140 μm, and 180 μm. (f)–(j) The TIE analysis corresponding to the Fresnel images in (a)–(e). The inset in the lower left corner of figures (f), (h), and (j) shows magnified views of the regions enclosed by the white dashed boxes, with white arrows indicating the direction of the helicity of the identified skyrmions. The colorwheel represents the in-plane magnetizations. (k) The contrast intensity at the center of the yellow region in figure (a) as a function of defocus value. (l) Profiles of the magnetic phase shift extracted from the yellow rectangular regions in figures (a), (b), and (c) (corresponding to defocus values of 20 μm, 60 μm, and 100 μm, respectively).

During the characterization of 15-nm skyrmion lattice using Fresnel imaging at various defocus values, we observed a significant reversal in magnetic contrast as a function of defocus. The TIE solution process requires the introduction of a non-zero constant, $q_0$, to prevent numerical divergence[48]; however, its value necessitates careful determination. In this study, a value of $q_0 = 9\times10^{-6}$ was adopted for the TIE reconstruction. Systematic comparative analysis confirmed this value to be optimal (see Figure S2), and it does not compromise the reliability of the conclusions drawn. As shown in **Figure 3**a, at a defocus value of 20 μm, the skyrmion exhibits bright contrast (white spots), and the corresponding TIE analysis (Figure 3f) reveals a CW helicity. When the defocus value is increased to 60 μm (Figure 3b), the bright contrast becomes very faint, which is also reflected in the TIE analysis (Figure 3g). Further increasing the defocus to 100 μm (Figure 3c) results in a complete transformation from bright to dark contrast (black spots), with the TIE analysis (Figure 3h) indicating an ACW helicity. We define this transition from bright to dark (or dark to bright) contrast as contrast reversal, and the corresponding difference in defocus values is denoted as $\Delta f$, where $\Delta f$ = 80 μm in this case. Upon further increasing the defocus to 180 μm (Figure 3e), the contrast of the skyrmion reverts from dark to bright, as evidenced by the TIE analysis (Figure 3j), showing that the helicity returns to a CW direction. The intermediate states during this transition, depicted in Figure 3d and 3i, exhibit almost no discernible magnetic contrast.

To quantitatively evaluate the variation in magnetic contrast intensity of 15-nm skyrmion lattice under different defocus conditions, we have plotted Figure 3k, which illustrates the trend of contrast intensity at the skyrmion center as a function of defocus value. The figure delineates the periodic transformation process experienced by the contrast intensity at the skyrmion center, transitioning from bright to dark, then bright again, and finally back to dark. Figure 3l presents the magnetic phase shift profiles extracted from the yellow rectangular regions in Figure 3a-3c, depicting how the magnetic phase shift within an entire skyrmion area varies with position under varying degrees of defocus (20, 60 and 100 μm).

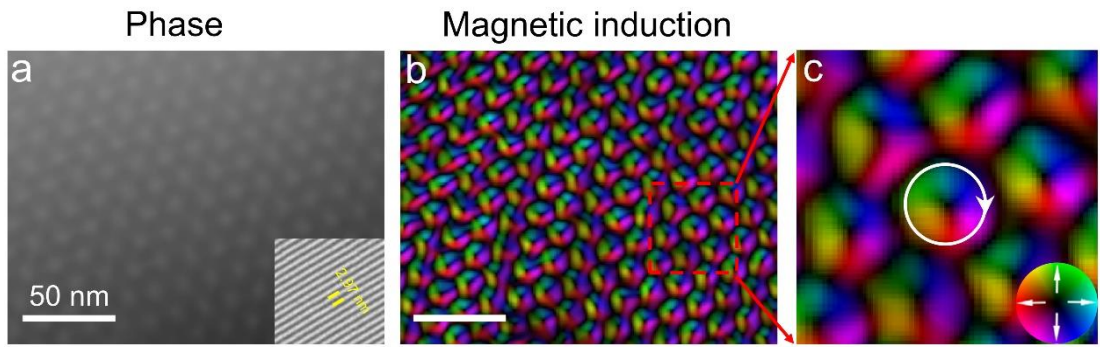

**Figure 4.** (a) Reconstructed magnetic phase obtained from electron holography. The inset on the right shows a magnified image of holographic interference fringes with a spacing of 2.97 nm. (b) Magnetic induction derived from the analysis of (a). (c) Enlargement of the region enclosed by the red dashed line in (b), where white arrows indicate the helicity of the skyrmions. The colorwheel represents the in-plane magnetizations.

However, in practice, simply adjusting the defocus value in Fresnel mode does not achieve periodic reversals of the skyrmion chirality. To more accurately characterize the genuine skyrmions in $Mn_{0.5}Fe_{0.5}Ge$, we employed off-axis electron holography (details in Experimental Section). This technique operates under in-focus conditions to avoid image delocalization that can occur in Lorentz mode. Figure 4a shows the reconstructed magnetic phase image obtained through off-axis electron holography, and further analysis of this image yielded the magnetic induction distribution (see Figure 4b). As illustrated in Figure 4c, the results demonstrate that the skyrmion helicity is CW.

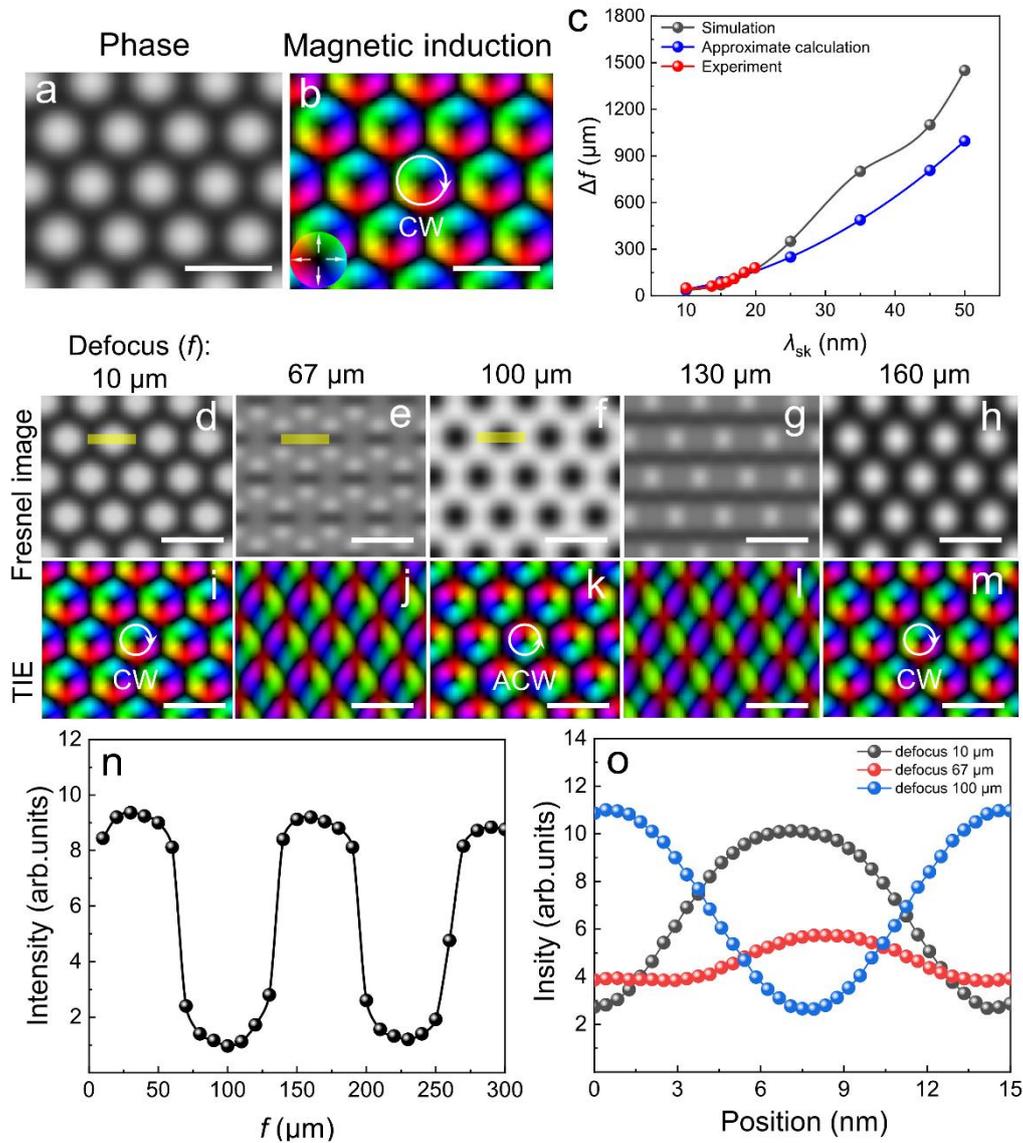

**Figure 5.** Numerical simulations of Fresnel imaging for ultrasmall skyrmions. (a)

simulated magnetic phase. (b) Magnetic induction derived from the analysis of (a). (c) The dependence of the $\Delta f$ on skyrmion size ($\lambda_{sk}$). (d)–(h), Simulated Fresnel images obtained at defocus values of 10 μm, 67 μm, 100 μm, 130 μm, and 160 μm. (i)–(m) The TIE analysis corresponds to the Fresnel images in (d)–(h), with white arrows indicating the direction of the helicity of the identified skyrmions. (n) The contrast intensity at the center of the yellow region in figure (d) as a function of defocus value. (o) Profiles of the magnetic phase shift extracted from the yellow rectangular regions in figures (d), (e), and (f) (corresponding to defocus values of 10 μm, 67 μm, and 100 μm, respectively). The colorwheel represents the in-plane magnetizations. The scale bar, 20 nm.

The contrast inversion phenomenon observed in Fresnel imaging of 15-nm skyrmion lattice has been perfectly reproduced in numerical simulations, as illustrated in **Figure 5**. Figure 5a shows the simulated magnetic phase image, while Figure 5b presents the magnetic induction derived from further analysis. In the simulated Fresnel mode, the skyrmion contrast exhibits periodic reversals as a function of defocus, as shown in Figure 5d-5m. Similarly, we have quantitatively evaluated the variations in magnetic contrast intensity under different defocus conditions, with the results plotted in Figure 5n and 5o. These simulation results are in excellent agreement with experimental observations. Notably, for non-periodic individual skyrmions, changes in defocus only affect the overall brightness and sharpness without inducing contrast reversal, as shown in Figure S3.

To further investigate the impact of skyrmion size on $\Delta f$ in Fresnel imaging, we conducted a systematic study. Initially, as illustrated in Figure S4–S6, the results indicate that neither sample thickness nor magnetic field strength significantly affects $\Delta f$. Furthermore, when the sign of the DMI is reversed, although the chirality of the corresponding skyrmion lattice is altered, the critical defocus distance $\Delta f$ associated with contrast reversal remains unchanged (Figure S7). Therefore, in our numerical simulations, we varied only the skyrmion size ($\lambda_{sk}$) parameter. The simulation results show a rapid increase in $\Delta f$ as the $\lambda_{sk}$ increases (Figure 5c), the discrepancy between the simulated and approximatively calculated curves primarily stems from the simplified model's approximation of the skyrmion as a single-frequency structure ($q \approx \frac{1}{\lambda_{sk}}$), whereas the actual imaging contrast in periodic skyrmion lattices arises from multi-frequency interference[49], leading to deviations in the predicted $\Delta f$ values. Despite these deviations, the simplified model still captures the overall trend of increasing $\Delta f$ with skyrmion size ($\lambda_{sk}$). Specifically, when the skyrmion size $\lambda_{sk}$ is 50 nm, the corresponding $\Delta f$ reaches 1450 μm (Figure S8). Moreover, by leveraging temperature to modulate the skyrmion size (Figure S9), we systematically acquired contrast reversal data across lattices of different dimensions. A clear trend of increasing $\Delta f$ with $\lambda_{sk}$ was observed (Figure 5c), consistent with our simulation results, thereby validating the predictive reliability of our method for skyrmions of varying sizes. Notably, the skyrmion lattice in FeGe ($\lambda_{sk} \approx 70$ nm) also exhibits contrast inversion, with the corresponding $\Delta f$ value reaching up to approximately 3000 μm, as shown in Figure S10. Based on these findings, as shown in Figure S11, we established two regions on the same sample, with skyrmions sized at 10 nm and 15 nm, respectively.

Due to the differences in $\Delta f$ between these two regions, within a specific defocus range, it is possible to observe skyrmions with contrasting black-and-white contrast on the same sample simultaneously. TIE analysis suggests that the helicity of the skyrmions in these two regions appears opposite; however, in reality, the helicity of both types of skyrmions is consistent.

**2.3 The physical mechanism of contrast reversal in imaging ultrasmall-sized skyrmion lattice in Lorentz-TEM Fresnel mode**

In TEM, the formation of images is influenced by the contrast transfer function (CTF)[50-53], which describes how phase information from the sample is translated into contrast variations in the final image. For Lorentz-TEM, the CTF can be expressed as follows:

$$CTF(q,f) = \sin(\pi\lambda q^2 f + \phi_0) \quad (2)$$

Here, $q$ represents the spatial frequency, $f$ denotes the defocus, $\lambda$ is the electron wavelength, and $\phi_0$ is the initial phase angle. This equation elucidates the defocus-dependent modulation of phase components across spatial frequencies, while incorporating the constraining effects of spatial coherence on electron wavefront phase distributions.

When characterizing periodic skyrmion lattices, the repetitive magnetic structure units within the lattice produce a series of diffraction spots in Fourier space[54], each corresponding to a specific spatial frequency $q$. Variations in defocus $f$ induce electron disturbance-mediated phase retardation, causing specific diffraction spots to traverse zero-crossings or extrema of the sinusoidal CTF. This results in periodic intensity fluctuations in real space, termed "contrast reversal." Such phenomena arise from defocus-altered interference conditions between diffraction orders.

To approximate the local magnetic structure of a skyrmion as a circular structure, the spatial frequency $q$ can be approximated as the inverse of the characteristic dimension $\lambda_{sk}$:

$$q \approx \frac{1}{\lambda_{sk}} \quad (3)$$

For a given spatial frequency $q$, to observe contrast reversal, where the CTF ($q$, $f$) changes sign from positive to negative or vice versa, specific defocus values $f$ must be satisfied. According to the CTF expression, contrast reversals occur when $\pi\lambda q^2 f$ equals $\frac{\pi}{2}$, $\frac{3\pi}{2}$, $\frac{5\pi}{2}$, etc. Therefore, the relationship between defocus $f$ and spatial frequency $q$ can be expressed as:

$$\pi\lambda q^2 f = (n + \frac{1}{2})\pi \quad (4)$$

where $n$ is an integer (0, 1, 2, ...). Simplifying this equation yields:

$$f = \frac{(2n+1)}{2\lambda q^2} \approx \frac{(2n+1)\lambda_{sk}^2}{2\lambda} \quad (5)$$

For larger skyrmions, achieving equivalent phase shifts necessitates greater defocus adjustments due to cumulative electron disturbance under spatial coherence constraints.

Individual skyrmions lack periodic arrangements, precluding discrete diffraction patterns. In Lorentz-TEM, their imaging relies on intrinsic topological magnetic structures and local field characteristics rather than defocus-modulated interference. Although defocus influences overall brightness, spatial coherence-preserved phase stability prevents complex contrast reversal. Single-skyrmion imaging thus depends on direct phase contrast rather than electron disturbance-driven interference effects.

Consequently, ultra-small skyrmion lattices, due to their periodicity, exhibit highly sensitive contrast changes in response to defocus variations in Lorentz-TEM, manifesting as periodic contrast reversals. Conversely, individual skyrmions, lacking such periodicity, do not display similar contrast reversal behavior. Moreover, as skyrmion size increases and spatial frequency decreases, achieving equivalent contrast reversal requires a greater defocus. It is noteworthy that the TIE is generally recommended for use under small-defocus conditions[55]. However, for ultrasmall skyrmions with $\lambda_{sk} \approx 15$ nm, the applicable upper limit of "small defocus" (approximately 15 μm, Figure S12) is significantly lower than the defocus values commonly used to characterize larger skyrmions[6], [56]. As a result, the conventionally understood "small defocus" condition may exceed the appropriate range for such dimensions, thereby inducing contrast reversal. Relying solely on black-white contrast to determine chirality can thus lead to misleading conclusions.

## 3. Conclusion

This study systematically investigates the ultrasmall 10-nm skyrmion lattice in B20-structured $Mn_{0.5}Fe_{0.5}Ge$ using Fresnel imaging in Lorentz-TEM. We observe periodic contrast reversals in magnetic images as a function of defocus, which are strongly correlated with the size of the skyrmions. Relying solely on black-and-white contrast for chirality determination may lead to misinterpretation. Through off-axis electron holography and numerical simulations, we elucidate how defocus-induced contrast changes impact chirality assessment. The simulations successfully reproduce the experimental contrast inversion, validating our findings. Furthermore, we demonstrate that the periodic reversal of magnetic contrast modulated by defocus arises from the sinusoidal modulation mechanism of the CTF in Lorentz-TEM. This highlights not only the influence of skyrmion size on imaging but also underscores the importance of understanding the role of defocus in Lorentz-TEM. The integration of multi-scale imaging techniques emerges as a prerequisite for fully unlocking the potential of ultra-small skyrmions in ultrahigh-density spintronic applications.

## 4. Experimental Section

*Sample preparation*: The initial step involved synthesizing polycrystalline $Mn_{0.5}Fe_{0.5}Ge$ by arc melting Mn granules (99.9% purity), Fe granules (99.9% purity), and Ge granules (99.999% purity) in their stoichiometric proportions. Subsequently, the material underwent high-pressure synthesis at 8 GPa and 1000°C for 2.5 hours, followed by a slow cooling process to room temperature.

*Preparation of $Mn_{0.5}Fe_{0.5}Ge$ lamellas*: The $Mn_{0.5}Fe_{0.5}Ge$ nanostructured lamellas were fabricated from a bulk single crystal using a standard lift-out method with an SEM-FIB dual-beam system (Helios Nanolab 600i, FEI).

*Lorentz-TEM measurements*: In-situ Fresnel imaging was performed in a Lorentz-TEM (Talos F200X, FEI) with an acceleration voltage of 200 kV. The TEM holder (model 636.6, Gatan) supported variable temperature measurements.

*Experimental details of off-axis Electron Holography*: Magnetic induction maps were acquired by using in situ off-axis electron holography under in-focus conditions to avoid image delocalization in Lorentz mode. Electron holography was performed using a single electrostatic biprism and a direct electron counting camera with an interference fringe spacing of 2.97 nm. Each experiment involved the acquisition of 30 object holograms, followed by 30 vacuum reference holograms to remove image distortions associated with the imaging and recording systems of the microscope. Averaging of the holograms, which were each acquired using an exposure time of 2 s, was used to improve the signal-to-noise ratio.

*Micromagnetic simulations*: The zero-temperature micromagnetic simulations for magnetic domains in 150-nm-thick lamellas were performed using MuMax3[57]. The simulated ratio *A/D* was chosen to be consistent with the skyrmion size varying from 10 nm to 70 nm according to the relation $\lambda_{sk} = 4\pi A/D$. The cell size adopted for the simulations was 0.5 nm × 0.5 nm × 5 nm.

*Simulation of Lorentz-TEM images and magnetic phase shift*: Lorentz-TEM images and magnetic phase images were simulated using the open-source software Pylorentz[58]. The spin configurations for these simulations were derived from micromagnetic simulation outputs. The simulation parameters included an accelerating voltage of 200 kV and defocus values. To evaluate the potential effects of microscope aberrations on Fresnel image simulations, supplementary simulations incorporating non-zero aberrations—specifically spherical aberration ($C_s$) and chromatic aberration ($C_c$)—were performed (Figure S13). Given that our experimental data were obtained using a Talos F200X TEM, which inherently exhibits non-zero $C_s$, the aberration parameters in our simulations were accordingly set to the default values in the Pylorentz software: $C_s = 2 \times 10^5$ nm and $C_c = 5 \times 10^6$ nm.

## Supporting Information

Supporting information is available from the Wiley Online Library or from the author.

## Ackonwledgments

This work was supported by the National Key R&D Program of China, Grant No. 2024YFA1611303 (J.T.); the Natural Science Foundation of China, Grants No. 521300103 (S.W.), 12422403 (J.T.), 52501305 (Y.Z.), U24A6001 (J.T.), 52325105 (H.D.), 12241406 (H.D.), 12174396 (J.T.), 12104123 (Y.W.), and 12204006 (W.L.); the Anhui Provincial Natural Science Foundation, Grant No. 2308085Y32 (J.T.), 2408085QA022 (Y.Z.), and 2508085MA018 (Y.W.); the Natural Science Project of Colleges and Universities in Anhui Province, Grants No. 2022AH030011 (J.T.) and 2024AH030046 (J.T.); CAS Project for Young Scientists in Basic Research, Grant No. YSBR-084 (H.D.); Systematic Fundamental Research Program Leveraging Major Scientific and Technological Infrastructure, Chinese Academy of Sciences, Grant No. JZHKYPT-2021-08 (H.D.); Anhui Province Excellent Young Teacher Training Project, Grant No. YQZD2023067 (Y.W.); the 2024 Project of GDRCYY No. 217 (Y.W.); the China Postdoctoral Science Foundation, Grant No. 2023M743543 (Y.W.) and 2024M760006 (Y.Z.); the Postdoctoral Fellowship Program of CPSF, Grant No.GZB20250792 (Y.Z.); the China Postdoctoral Science Foundation - Anhui Joint Support Program, Grant No. 2025T003AH (Y.Z.); Anhui Postdoctoral Scientific Research Program Foundation, Grant No. 2025B1080 (Y.Z.).

## Conflict of Interest

## Author Contributions

Y.Z. and W.L. contributed equally to this work. S.W., H.D., M.T., and J.T. supervised the project. J.T., and Y.Z. conceived the idea and designed the experiments. W.L. synthesized the bulk crystals. Y.Z., H.Z., and Y.W. fabricated the $Mn_{0.5}Fe_{0.5}Ge$ lamella and performed the TEM measurements with the help of J.J.. M.S. and S.Q. performed the simulations. Y.Z., Y.W., and J.T. wrote the manuscript with input from all authors. All authors discussed the results and contributed to the manuscript.

## Data Availability Statement

The data that support the findings of this study are available from the corresponding author upon reasonable request.

## References

[1]  S. Mühlbauer, B. Binz, F. Jonietz, C. Pfleiderer, A. Rosch, A. Neubauer, R. Georgii, P. Böni, Skyrmion lattice in a chiral magnet. *Science* **2009**, *323* (5916), 915-919.

[2]  I. Kézsmárki, S. Bordács, P. Milde, E. Neuber, L. M. Eng, J. S. White, H. M. Rønnow, C. D. Dewhurst, M. Mochizuki, K. Yanai, H. Nakamura, D. Ehlers, V. Tsurkan, A. Loidl, Néel-type skyrmion lattice with confined orientation in the polar magnetic semiconductor $GaV_4S_8$. *Nat. Mater.* **2015**, *14* (11), 1116-1122.

[3]  S. Heinze, K. von Bergmann, M. Menzel, J. Brede, A. Kubetzka, R. Wiesendanger, G. Bihlmayer, S. Blügel, Spontaneous atomic-scale magnetic skyrmion lattice in two dimensions. *Nat. Phys.* **2011**, *7* (9), 713-718.

[4]  D. Song, W. Wang, S. Zhang, Y. Liu, N. Wang, F. Zheng, M. Tian, R. E. Dunin-Borkowski, J. Zang, H. Du, Steady motion of 80-nm-size skyrmions in a 100-nm-wide track. *Nat. Commun* **2024**, *15* (1), 5614.

[5] K. Wang, V. Bheemarasetty, J. Duan, S. Zhou, G. Xiao, Fundamental physics and applications of skyrmions: A review. *J MAGN MAGN MATER* **2022**, *563*, 169905.

[6] J. Tang, Y. Wu, W. Wang, L. Kong, B. Lv, W. Wei, J. Zang, M. Tian, H. Du, Magnetic skyrmion bundles and their current-driven dynamics. *Nat. Nanotechnol* **2021**, *16* (10), 1086-1091.

[7] S. Woo, K. M. Song, H.-S. Han, M.-S. Jung, M.-Y. Im, K.-S. Lee, K. S. Song, P. Fischer, J.-I. Hong, J. W. Choi, B.-C. Min, H. C. Koo, J. Chang, Spin-orbit torque-driven skyrmion dynamics revealed by time-resolved X-ray microscopy. *Nat. Commun* **2017**, *8* (1), 15573.

[8] H. Shi, J. Zhang, Y. Xi, H. Li, J. Chen, I. Ahmed, Z. Ma, N. Cheng, X. Zhou, H. Jin, X. Zhou, J. Liu, Y. Sun, J. Wang, J. Li, T. Yu, W. Hao, S. Zhang, Y. Du, Dynamic behavior of above-room-temperature robust skyrmions in 2D van der Waals magnet. *Nano Lett.* **2024**, *24* (36), 11246-11254.

[9] J. Yang, J. Kim, C. Abert, D. Suess, S.-K. Kim, Stability of skyrmion formation and its abnormal dynamic modes in magnetic nanotubes. *Phys. Rev. B* **2020**, *102* (9), 094439.

[10] R. Blasing, A. A. Khan, P. Filippou, C. Garg, F. Hameed, J. Castrillón, S. Parkin, Magnetic racetrack memory: from physics to the cusp of applications within a decade. *Proc. IEEE* **2020**, *PP*, 1-19.

[11] S. Luo, L. You, Skyrmion devices for memory and logic applications. *APL Materials* **2021**, *9* (5), 050901.

[12] K. Ishibashi, S. Yorozu, T. Arima, M. Kawamura, Y. Tokura, K. Karube, X. Yu, Y. Taguchi, T. Hanaguri, T. Machida, Y. M. Itahashi, Y. Iwasa, H. Nishikawa, F. Araoka, T. Hioki, E. Saitoh, R. S. Deacon, M. Yamamoto, N. Fang, Y. K. Kato, A. Hida, M. Takamoto, H. Katori, S. de Léséleuc, T. Aoki, H. Yonezawa, A. Furusawa, Y. Tabuchi, S. Tamate, E. Abe, Y. Nakamura, T. Nakajima, S. Tarucha, K. Seki, T. Shirakawa, S. Yunoki, N. Nagaosa, Research on Quantum Materials and Quantum Technology at RIKEN. *ACS Nano* **2025**, *19*, 12427-12457.

[13] A. Fert, N. Reyren, V. Cros, Magnetic skyrmions: advances in physics and potential

applications. *Nat. Rev. Mater.* **2017**, *2* (7), 17031.

[14] A. N. Bogdanov, C. Panagopoulos, Physical foundations and basic properties of magnetic skyrmions. *Nat. Rev. Phys.* **2020**, *2* (9), 492-498.

[15] A. K. Nayak, V. Kumar, T. Ma, P. Werner, E. Pippel, R. Sahoo, F. Damay, U. K. Rößler, C. Felser, S. S. P. Parkin, Magnetic antiskyrmions above room temperature in tetragonal Heusler materials. *Nature* **2017**, *548* (7669), 561-566.

[16] X. Yu, N. Kanazawa, Y. Onose, K. Kimoto, W. Z. Zhang, S. Ishiwata, Y. Matsui, Y. Tokura, Near room-temperature formation of a skyrmion crystal in thin-films of the helimagnet FeGe. *Nat. Mater.* **2011**, *10* (2), 106-109.

[17] M. T. Birch, L. Powalla, S. Wintz, O. Hovorka, K. Litzius, J. C. Loudon, L. A. Turnbull, V. Nehruji, K. Son, C. Bubeck, T. G. Rauch, M. Weigand, E. Goering, M. Burghard, G. Schütz, History-dependent domain and skyrmion formation in 2D van der Waals magnet $Fe_3GeTe_2$. *Nat. Commun* **2022**, *13* (1), 3035.

[18] H. Vakili, J.-W. Xu, W. Zhou, M. N. Sakib, M. G. Morshed, T. Hartnett, Y. Quessab, K. Litzius, C. T. Ma, S. Ganguly, M. R. Stan, P. V. Balachandran, G. S. D. Beach, S. J. Poon, A. D. Kent, A. W. Ghosh, Skyrmionics—Computing and memory technologies based on topological excitations in magnets. *J. Appl. Phys.* **2021**, *130* (7), 070908.

[19] B. Dieny, I. L. Prejbeanu, K. Garello, P. Gambardella, P. Freitas, R. Lehndorff, W. Raberg, U. Ebels, S. O. Demokritov, J. Akerman, A. Deac, P. Pirro, C. Adelmann, A. Anane, A. V. Chumak, A. Hirohata, S. Mangin, S. O. Valenzuela, M. C. Onbaşlı, M. d'Aquino, G. Prenat, G. Finocchio, L. Lopez-Diaz, R. Chantrell, O. Chubykalo-Fesenko, P. Bortolotti, Opportunities and challenges for spintronics in the microelectronics industry. *Nat. Electron* **2020**, *3* (8), 446-459.

[20] H. Zhang, D. Zhu, W. Kang, Y. Zhang, W. Zhao, Stochastic computing implemented by skyrmionic logic devices. *Phys. Rev. Appl* **2020**, *13*, 054049.

[21] K. Palotás, L. Rózsa, E. Simon, L. Udvardi, L. Szunyogh, Spin-polarized scanning tunneling microscopy characteristics of skyrmionic spin structures exhibiting various topologies. *Phys. Rev. B* **2017**, *96* (2), 024410.

[22] X. Yu, Y. Onose, N. Kanazawa, J. H. Park, J. H. Han, Y. Matsui, N. Nagaosa, Y.

Tokura, Real-space observation of a two-dimensional skyrmion crystal. *Nature* **2010**, *465* (7300), 901-904.

[23] K. Shibata, X. Z. Yu, T. Hara, D. Morikawa, N. Kanazawa, K. Kimoto, S. Ishiwata, Y. Matsui, Y. Tokura, Towards control of the size and helicity of skyrmions in helimagnetic alloys by spin–orbit coupling. *Nat. Nanotechnol* **2013**, *8* (10), 723-728.

[24] L. Peng, Y. Zhang, S.-l. Zuo, M. He, J.-w. Cai, S.-g. Wang, H.-x. Wei, J.-q. Li, T.-y. Zhao, B.-g. Shen, Lorentz transmission electron microscopy studies on topological magnetic domains. *Chin. Phys. B* **2018**, *27* (6), 066802.

[25] M. Tanase, A. K. Petford-Long, In situ TEM observation of magnetic materials. *MICROSC RES TECHNIQ* **2009**, *72* (3), 187-196.

[26] W. Wang, D. Song, W. Wei, P. Nan, S. Zhang, B. Ge, M. Tian, J. Zang, H. Du, Electrical manipulation of skyrmions in a chiral magnet. *Nat. Commun* **2022**, *13* (1), 1593.

[27] J. Tang, J. Jiang, Y. Wu, L. Kong, Y. Wang, J. Li, Y. Soh, Y. Xiong, S. Wang, M. Tian, H. Du, Creating and deleting a single dipolar skyrmion by surface spin twists. *Nano Lett.* **2025**, *25* (1), 121-128.

[28] J. Tang, J. Jiang, N. Wang, Y. Wu, Y. Wang, J. Li, Y. Soh, Y. Xiong, L. Kong, S. Wang, M. Tian, H. Du, Combined magnetic imaging and anisotropic magnetoresistance detection of dipolar skyrmions. *Adv. Funct. Mater.* **2023**, *33* (4), 2207770.

[29] X. Yu, D. Morikawa, T. Yokouchi, K. Shibata, N. Kanazawa, F. Kagawa, T.-h. Arima, Y. Tokura, Aggregation and collapse dynamics of skyrmions in a non-equilibrium state. *Nat. Phys.* **2018**, *14* (8), 832-836.

[30] X. Yu, F. Kagawa, S. Seki, M. Kubota, J. Masell, F. S. Yasin, K. Nakajima, M. Nakamura, M. Kawasaki, N. Nagaosa, Y. Tokura, Real-space observations of 60-nm skyrmion dynamics in an insulating magnet under low heat flow. *Nat. Commun* **2021**, *12* (1), 5079.

[31] H. S. Park, X. Yu, S. Aizawa, T. Tanigaki, T. Akashi, Y. Takahashi, T. Matsuda, N. Kanazawa, Y. Onose, D. Shindo, A. Tonomura, Y. Tokura, Observation of the magnetic flux and three-dimensional structure of skyrmion lattices by electron holography. *Nat.*

*Nanotechnol* **2014**, *9* (5), 337-342.

[32] P. A. Midgley, An introduction to off-axis electron holography. *MICRON* **2001**, *32* (2), 167-184.

[33] P. A. Midgley, R. E. Dunin-Borkowski, Electron tomography and holography in materials science. *Nat. Mater.* **2009**, *8* (4), 271-280.

[34] J. C. Loudon, A. C. Twitchett-Harrison, D. Cortés-Ortuño, M. T. Birch, L. A. Turnbull, A. Štefančič, F. Y. Ogrin, E. O. Burgos-Parra, N. Bukin, A. Laurenson, H. Popescu, M. Beg, O. Hovorka, H. Fangohr, P. A. Midgley, G. Balakrishnan, P. D. Hatton, Do images of biskyrmions show type-II bubbles? *Adv. Mater.* **2019**, *31* (16), 1806598.

[35] J. Tang, Y. Wu, L. Kong, W. Wang, Y. Chen, Y. Wang, Y. Soh, Y. Xiong, M. Tian, H. Du, Two-dimensional characterization of three-dimensional magnetic bubbles in $Fe_3Sn_2$ nanostructures. *Natl. Sci. Rev.* **2021**, *8* (6), nwaa200.

[36] Y. Chen, B. Lv, Y. Wu, Q. Hu, J. Li, Y. Wang, Y. Xiong, J. Gao, J. Tang, M. Tian, H. Du, Effects of tilted magnetocrystalline anisotropy on magnetic domains in $Fe_3Sn_2$ thin plates. *Phys. Rev. B* **2021**, *103* (21), 214435.

[37] Y. Yao, B. Ding, J. Cui, X. Shen, Y. Wang, W. Wang, R. Yu, Magnetic hard nanobubble: A possible magnetization structure behind the bi-skyrmion. *Appl. Phys. Lett.* **2019**, *114* (10).

[38] J. Tang, Y. Wu, J. Jiang, L. Kong, W. Liu, S. Wang, M. Tian, H. Du, Sewing skyrmion and antiskyrmion by quadrupole of Bloch points. *Sci. Bull.* **2023**, *68*, 2919-2923.

[39] S. V. Grigoriev, N. M. Potapova, S. A. Siegfried, V. A. Dyadkin, E. V. Moskvin, V. Dmitriev, D. Menzel, C. D. Dewhurst, D. Chernyshov, R. A. Sadykov, L. N. Fomicheva, A. V. Tsvyashchenko, Chiral properties of structure and magnetism in $Mn_{1-x}Fe_xGe$ compounds: when the left and the right are fighting, who wins? *Phys. Rev. Lett.* **2013**, *110* (20), 207201.

[40] J. P. Chen, Y. L. Xie, Z. B. Yan, J.-M. Liu, Tunable magnetic helicity in $Mn_{1-x}Fe_xGe$: A Monte Carlo simulation. *J. Appl. Phys.* **2015**, *117* (17), 17C750.

[41] T. Tanigaki, K. Shibata, N. Kanazawa, X. Yu, Y. Onose, H. S. Park, D. Shindo, Y.

Tokura, Real-Space Observation of Short-Period Cubic Lattice of Skyrmions in MnGe. *Nano Lett.* **2015**, *15* (8), 5438-5442.

[42] L. D. Landau, E. M. Lifshitz, in (Eds.: L. D. Landau, E. M. Lifshitz), Pergamon, Amsterdam **1984**.

[43] Y. Yao, B. Ding, J. Liang, H. Li, X. Shen, R. Yu, W. Wang, Chirality flips of skyrmion bubbles. *Nat. Commun* **2022**, *13* (1), 5991.

[44] C.-E. Fillion, J. Fischer, R. Kumar, A. Fassatoui, S. Pizzini, L. Ranno, D. Ourdani, M. Belmeguenai, Y. Roussigné, S.-M. Chérif, S. Auffret, I. Joumard, O. Boulle, G. Gaudin, L. Buda-Prejbeanu, C. Baraduc, H. Béa, Gate-controlled skyrmion and domain wall chirality. *Nat. Commun* **2022**, *13* (1), 5257.

[45] S. Seki, X. Yu, S. Ishiwata, Y. Tokura, Observation of skyrmions in a multiferroic material. *Science* **2012**, *336* (6078), 198-201.

[46] Y. Zhang, J. Tang, Y. Wu, M. Shi, X. Xu, S. Wang, M. Tian, H. Du, Stable skyrmion bundles at room temperature and zero magnetic field in a chiral magnet. *Nat. Commun* **2024**, *15* (1), 3391.

[47] X. Zhao, J. Tang, K. Pei, W. Wang, S.-Z. Lin, H. Du, M. Tian, R. Che, Current-induced magnetic skyrmions with controllable polarities in the helical phase. *Nano Lett.* **2022**, *22* (22), 8793-8800.

[48] J. Cui, Y. Yao, X. Shen, Y. G. Wang, R. C. Yu, Artifacts in magnetic spirals retrieved by transport of intensity equation (TIE). *J MAGN MAGN MATER* **2018**, *454*, 304-313.

[49] J. N. Chapman, The investigation of magnetic domain structures in thin foils by electron microscopy. *J PHYS D APPL PHYS* **1984**, *17* (4), 623.

[50] Y. Cong, S. J. Ludtke, in (Ed.: G. J. Jensen), Academic Press, **2010**.

[51] L. K. Sheth, A. L. Piotrowski, N. R. Voss, Visualization and quality assessment of the contrast transfer function estimation. *J. Struct. Biol.* **2015**, *192* (2), 222-234.

[52] N. Tanaka, in (Ed.: N. Tanaka), Springer Japan, Tokyo **2017**.

[53] C. Phatak, A. K. Petford-Long, M. De Graef, Recent advances in Lorentz microscopy. *CURR OPIN SOLID ST M* **2016**, *20* (2), 107-114.

[54] E. Balkind, A. Isidori, M. Eschrig, Magnetic skyrmion lattice by the Fourier

transform method. *Phys. Rev. B* **2019**, *99* (13), 134446.

[55] M. Beleggia, M. A. Schofield, V. V. Volkov, Y. Zhu, On the transport of intensity technique for phase retrieval. *Ultramicroscopy* **2004**, *102* (1), 37-49.

[56] X. Z. Yu, N. Kanazawa, Y. Onose, K. Kimoto, W. Z. Zhang, S. Ishiwata, Y. Matsui, Y. Tokura, Near room-temperature formation of a skyrmion crystal in thin-films of the helimagnet FeGe. *Nat. Mater.* **2011**, *10* (2), 106-109.

[57] A. Vansteenkiste, J. Leliaert, M. Dvornik, M. Helsen, F. Garcia-Sanchez, B. Van Waeyenberge, The design and verification of MuMax3. *AIP Advances* **2014**, *4* (10), 107133.

[58] A. R. C. McCray, T. Cote, Y. Li, A. K. Petford-Long, C. Phatak, Understanding complex magnetic spin textures with simulation-assisted Lorentz transmission electron microscopy. *Phys. Rev. Appl* **2021**, *15* (4), 044025.